\documentstyle{article}
  \advance\oddsidemargin  by -1in
  \advance\evensidemargin by -1in
  \advance\topmargin      by -0.5in
\def\lsim{\mathrel{\rlap{\lower4pt\hbox{\hskip1pt$\sim$}}
    \raise1pt\hbox{$<$}}}         
\def\gsim{\mathrel{\rlap{\lower4pt\hbox{\hskip1pt$\sim$}}
    \raise1pt\hbox{$>$}}}         

\def\q2{Q^2}
\def\beq{\begin{equation}}
\def\kt{k_{\perp}}
\def\ks{\kappa_{\perp}}

\def\endeq{\end{equation}}
\def\arr{\begin{eqnarray}}
\def\endarr{\end{eqnarray}}
\makeindex

\begin{document}

\large
\phantom.\hspace{9.2cm}{\Large\bf ITEP-12-95\bigskip\\}
\phantom.\hspace{9.2cm}{\Large \bf February 1995}\vspace{1.5cm}\\
\begin{center}
{\bf \huge
Exclusive high-$Q^2$ electroproduction:\\
light-cone wave functions\\
 and\\
electromagnetic form factors of mesons.
\vspace{1.0cm}}

{\Large J.Speth$^{a)}$ and
V.R.Zoller$^{b)}$\medskip\\ }
{\sl
$^{a)}$Institute f\"ur Kernphysik, Forshungszentrum J\"ulich,\\
D-52425 J\"ulich, Germany\medskip\\
$^{b)}$Institute for Theoretical and Experimental Physics,\\
ul. B.Chermushkinskaya 25, 117259 Moscow, Russia\\
E-Mail: ZOLLER@VXITEP.ITEP.RU}
\vspace{1cm}\\
{\bf           Abstract}
\end{center}
	In the light-cone technique, one can describe exclusive
 electroproduction treating mesons and baryons as (non-pointlike)
 partons  of the nucleon. In this technique, the non-perturbative light-cone
 wave function of the meson-baryon Fock state of the physical nucleon defines
 the universal, projectile independent, density (flux) of
 on-mass-shell mesons in the proton. We apply the light-cone
 technique  to electroproduction reaction $ep\to e\pi^+n$ long considered as
 a means  of measuring the electromagnetic form factor of the pion. We
 show that the interpretation  of the long-standing puzzle of large
transverse cross section ($\sigma_T$) in terms of the $\gamma^*\rho\to \pi$
 transition on the $\rho$ mesons in the light-cone proton is possible, but
 requires quite a slow decrease of the $F_{\rho\pi}(\q2)$ form factor. This
 interpretation can be tested in the related $ep\to e\pi^0p$
reaction.

	Corrections which are due to the final-state meson-baryon
interactions (FSI)
 are evaluated and are shown to amount to
  a $25\%$ effect at  moderately large
$\q2$.
  Vanishing FSI with increasing $Q^2$ - the
 color transparency
phenomenon -  is shown to be very strong.

\newpage



\section{Introduction}

	The data on pion electroproduction above the  resonance
 region are analyzed in terms of nucleon and pion pole diagrams, shown in
 Fig.1. If the kinematics are chosen appropriately one argues  that the
pion pole diagram dominates [1]. This observation
 is used in determining the pion electromagnetic form factor [2,3].
Also it has been predicted [2,3,4], that at high $Q^2$
 the total cross section $\sigma(ep\to\pi^+n)$ is dominated
by the longitudinal
(scalar) component ($\sigma_L$)
  and the ratio $\sigma_L/\sigma_T$ grows
 rapidly at high $Q^2$ reaching the value  $\sigma_L/\sigma_T\approx 15$
 already at $Q^2\approx 3\,GeV^2$. The analysis of data taken at small and
 large values of the polarization parameter made it possible to separate
 the longitudinal and transverse components of the electroproduction cross
 section for $Q^2=1.2-3.3\,GeV^2$ [2]. Surprisingly, the measured ratio
 $\sigma_L/\sigma_T$ appeared to be  decreasing function of $Q^2$
 which takes the value  $\sigma_L/\sigma_T\sim 1$ at
$Q^2=3.3\, GeV^2$ in strong contradiction with the  theoretical
 estimates [2,3,4].

Anticipating the results, we
explain that the effect of strong absorption of the transverse photons is
entirely due to the admixture of the transversely polarized $\rho$
mesons in the light-cone proton and the final state pion is generated
in the magnetic dipole transition $\gamma^*\rho\to \pi$. The
longitudinal component of total cross section at high $Q^2$ is
determined by the light-cone density of the $\pi N$ Fock states in
the proton.

The
observables
in  exclusive electroproduction are very sensitive to the dynamics of the
momentum transfer in the $\pi NN$ vertex [5].  In the traditional approach
the uncorrelated monopole (dipole) $t$- and $u$-channel pion-nucleon form
factors are introduced.  Such a description has many flaws. In particular
it violates the electric charge conservation and do not satisfy the
energy-momentum sum rules. An attempt to minimize the momentum
non-conservation effects by the appropriate choice of the $\pi NN$ form
factor has been undertaken in [6].

On the way toward
 the consistent relativistic
description of the meson-baryon component of the nucleon
the  Llewellyn Smith's (LS) paper [7]
 was of the particular importance. In [7]
  the role of the energy-momentum sum rules
has been emphasized
and the LS-ansatz (see below) has been formulated as a
constraint on the functional form of the flux of pions and nucleons.
In [8,9]  the
light-cone wave function (LCWF) technique [10] was
extended  to the meson/baryon
exchange processes
and the light-cone densities of $\pi N$, $\pi \Delta$, $K\Lambda$,
 $\eta N$  Fock states have been derived.
 The corollary of
 the LCWF is that the LS-ansatz and the local gauge invariance,
 and momentum
conservation thereof, are satisfied by construction.

This  approach has been  applied
recently to the processes involving vector mesons.
In this case, as well as in the case of spin-$3/2$,
(corresponding Lagrangians involve the couplings with derivatives),
 some care is needed to
avoid the conflict with gauge invariance,
as it has been emphasized in [9].
 The  results consistent with gauge invariance were obtained in ref. [11].

The light-cone approach has one more important virtue
evidently related to those mentioned above.
Namely, it allows one to interpret the densities of mesons and baryons as
the  (non-perturbative) partons of the physical nucleon
 and, what is much more substantial, to study them
like  parton densities in inclusive DIS. The only difference is, however,
that the  exclusive final states have to be analysed in the
one-meson electroproduction with the
partonic kinematics.
Indeed, in this case
the photon of energy $\nu$ and virtuality $\q2$,
greatly exceeding the parton
(meson/baryon)  virtuality $k^2$,
\beq
 \q2\gg m^2,\,\,
  k_{\perp}^2\,,
\endeq
probes the density of
mesons and/or baryons in the proton at the light-cone Sudakov variable
\beq
\alpha\,=\,{\q2\over 2m_p\nu}\equiv\,x.
\endeq
It is important that the cross section of the reaction
$ep\to eMB$
does factorize
\beq
{d\sigma\over dxdQ^2}\sim f_{\pi N}(x)\,F^2_{\pi(N)}(\q2)\,.
\endeq
Here, $f_{\pi N}(\alpha)$ is the parton
 (meson/baryon) density, which scales, and
$F_{\pi(N)}(\q2)$
is
the on-mass-shell electromagnetic form factor of the struck parton.

Factorization implies a possibility of  separate
 analysis  of the light-cone meson-baryon
  density functions and the electromagnetic form
  factors  of mesons and baryons.
Consideration of the problems related to the practical realization
of such a program
 is the subject of the present  paper.

Previously, the pion
electroproduction at small $\q2$  has been studied in terms of a pion
distribution function in  [12].  For the perturbative QCD
consideration of the exclusive processes see ref.[13].




\section{Single pion electroproduction:
light-cone parton densities and form factors}

We find it convenient to use a light-cone momentum notation where a vector
is denoted by
$$
k_{\mu}=(k_-,k_+,\kt)=(\beta p_-,\alpha p_+,\kt)
$$
 with
$$
k_\pm=(k_0\pm k_3)/\sqrt{2}.
$$
 For the process
(4) we choose a frame where
$$
p_\mu=(p_-,p_+,0,0)
$$
and
$$q_\mu=(q_-,q_+,0,0)$$
with the Bjorken variable
$$x=\,-{q^2/ 2pq}=\alpha=
\,-{q_+/p_+}\left(1+O\left(m_N^2/Q^2\right)\right)\,.$$
That is we choose
$p_+q_-\gg p_-q_+$.
Still, it is assumed  that the invariant mass, $W$, of the final hadronic
state is large enough, $W^2\approx\q2(1-x)/x>5\, GeV^2$, so that we do not
encounter the problems
of the resonance physics [14].

The differential cross section of the exclusive reaction (Fig.1a)
\beq
ep\to e\pi^+n
\endeq
takes the form
\begin{eqnarray}
{d\sigma(
ep\to e\pi^+n)
\over dxdQ^2dk^2_{\perp}}=2K_L(x,\q2)
 f_{\pi N}(1-x,k^2_{\perp})
F_{\pi}^2(\q2)\nonumber\\
+
2K_T(x,\q2)f^T_{\rho N}(1-x,\kt^2)
{1\over 8}\q2
F^2_{\rho\pi}(\q2)
\end{eqnarray}
where
$f_{\pi N}(\alpha,\kt^2)$
 is the density of nucleons
with the transverse momentum
$\kt$ and the fraction $\alpha$ of the proton's light-cone momentum.
While
$f_{\pi N}(1-\alpha,\kt^2)$ represents
the density of pions with the momentum fraction $1-\alpha$, in
agreement with the LS-ansatz.

In other words $f_{\pi N}$,
being normalized to the number of nucleons (pions)
$n_{N(\pi)}$, represents the square of the entire
 LCWF
of the $\pi N$ Fock state
\beq
f_{\pi N}(\alpha,\kt^2)=|\phi_{\pi N}(\alpha,\vec\kt)|^2
\endeq
which involves not only radial but also the spin-orbit degrees
of freedom.
With the $\pi NN$ vertex of the form
$g_{N\pi N}\bar u\gamma_5 u\varphi_{\pi}\Gamma$
the density of $\pi^0 p$ states  reads [8,9]
\beq
f_{\pi N}(\alpha,\kt^2)={g^2_{p\pi^0p}\over 16\pi^2}
{\left[m^2_N(1-\alpha)^2+\kt^2\right]\over \alpha^2(1-\alpha)}
|\varphi(M^2_{\pi N}(\alpha,\kt^2))|^2.
\endeq
Hence, the isospin factors $2$ in rhs of eq.(5)

The radial part of the LCWF,
\beq
\varphi(M^2)={\Gamma(M^2)\over M^2-m_p^2}\,,
\endeq
depends only on the invariant mass squared
of the two-body meson-baryon (MB) state,
\beq
M^2_{MB}(\alpha,\kt^2)= {m_B^2+k^2_{\perp}\over  \alpha}+
{m_M^2+k^2_{\perp}\over 1-\alpha}\,.
\endeq
 The vertex function
$
\Gamma(M^2)
$ is parametrized as
\beq
\Gamma(M^2_{MB})=\exp\left[-{1\over 2}R^2_{MB}\left
(M^2_{MB}-M_0^2\right)\right].
\endeq
 Originally [8,9],  the $M NB$ couplings, $g_{MNB}$, as well as
still another nonperturbative parameter - the
radii $R_{MB}$ of the meson baryon Fock state
were inferred from
an analysis of the experimental data on the fragmentation of high-energy proton
into nucleons and hyperons -
the process dominated by stripping off the mesons of the meson-baryon
Fock states [15]. The normalization to the point
$M_0^2=(m_B+m_M)^2$
of the $MB$-system threshold
was accepted.
 The continuation into the unphysical region of
the meson pole, $M_0^2=m_N^2$, accepted in many of the
recent papers,
calls
for knowledge of the final state interaction effects which
are discussed below.

The  kinematical factors $K_L$ and $K_T$ in (5) are as follows
\beq
K_T(x,Q^2)= {4\pi\alpha_{em}^2\over Q^4}(1-y+{1\over 2}y^2)
\endeq
\beq
K_L(x,Q^2)= {4\pi\alpha_{em}^2\over Q^4}(1-y),\,\,
 y={\q2\over xs},\,\,s=2pp_e,
\endeq
where
 $p_e$ and $p$
are the four-momenta of the beam-electron and the target-proton,
respectively.

In (5) $F_{\pi}(\q2)$ is the on-shell charge form factor of the pion.
It is worth recalling that
in the light-cone  parton model the condition (1)
guarantees the on-shellness of partons [16].
 Still,  the dispersion integral representation [17]
which  deals only with the on-mass-shell internal  particles
in corresponding unitarity diagrams
reproduces the above factorization.
Note in this connection that the vertex function $\Gamma(M^2)$ (10)
cuts off the large transverse momenta at $\kt^2\sim \alpha(1-\alpha)/R^2$.
The typical $\alpha$'s we are dealing with are
\beq
0.2\lsim \alpha \lsim 0.8
\endeq
In this specific region the reggeization effects can be neglected.
 Then,
for $R^2\simeq 1\, GeV^{-2}$
the main contribution to the cross section (5)
comes from rather small transverse momenta $\kt^2\lsim 0.3\, GeV^{2}$.

The longitudinal momentum partition in the $\pi N-$ system, as it is prescribed
by the $\pi N$ LCWF, is very far from uniformity. The nucleon
carries the momentum fraction
\beq
\alpha\sim m_N/\left(m_N+\sqrt{m^2_{\pi}+\kt^2}\right),
\endeq
 only the rest,
 $1-\alpha$,
 falls to the pion. Such a disbalance  leads to a
suppression of the pion-pole contribution
to the cross section (5)
for $\alpha\gsim\,0.7$. By the same reasons the nucleon-pole contribution
vanishes for  $\alpha\lsim\,0.3$.


The second term in (5) is due to the $\rho^+-\pi^+$ transition which is
 under the control of the form factor
$F_{\rho\pi}(\q2)$
\beq
\langle\pi(k^{\prime})
|J^{\mu}_{el}(0)|
\rho(k)\rangle=
F_{\rho\pi}(\q2)\epsilon^{\mu\nu\lambda\sigma}
k^{\prime}_{\nu}k_{\lambda}\rho_{\sigma}\,,
\endeq
where $\rho_{\sigma}$ is the
$\rho$-meson
polarization vector.
Since the amplitude (15)
is purely transversal, it has been predicted  [18,19] that at
asymptotically high $\q2$
\beq
F_{\rho\pi}(\q2)\sim Q^{-4}\,.
\endeq
 At the same time, the magnetic dipole (M1) transition (15)
generates an extra power of $Q^2$ in the electroproduction cross section (5).

In the region of $\q2\sim 5-10\,GeV^2$, which we are interested in,
the form factor $F_{\rho\pi}$
is not calculable within pQCD
 (for more discussion on
  the slow onset of the hard scattering regime in the form factors
  see [20]). This is the reason why
we  resort to the parametrization
of $F_{\rho\pi}$ as
\beq
F_{\rho\pi}(\q2)={g_{\rho\pi\gamma}
\left(1+\q2/\Lambda_{\rho\pi}^2\right)^{-2}} \,,
\endeq
where $\Lambda_{\rho\pi}$ should be chosen to match the experimental data.
The normalization is such that
$F_{\rho\pi}(0)=
g_{\rho\pi\gamma}
$.
 The $SU(3)$ quark model prediction [21,22]
$g_{\rho\pi\gamma}=2/3\mu_p$, $\mu_p=2.79/
m_p$, which we rely upon, does not contradict the measured
$\rho^{\pm}$ radiative decay width [23].

The density of the transversely polarized
$\rho^0$-mesons in the proton, $f^T_{\rho N}$, can easily be obtained
\begin{eqnarray}
f^T_{\rho N}(\alpha,\kt^2)=
{g^2_{p\rho^0p}\over 16\pi^2}{m_N^2\over \alpha^2(1-\alpha)}\nonumber\\
\times\left\{2\left[(1-\alpha)^2+\tau(1+\alpha^2)\right]
+4(f/g)(1-\alpha)^2\right.\nonumber\\
+\left.(f/g)^2(1-\alpha)^2(\tau+2)(\tau+1)\right\}
\left|\varphi(M^2_{\rho N})\right|^2.
\end{eqnarray}
Here $g$ and $f$ are the vector and tensor $\rho NN$ couplings and
   the dimensionless variable $\tau$ is introduced
\beq
\tau={\kt^2\over m_N^2(1-\alpha)^2}\,.
\endeq
Our definition of the meson-baryon coupling constants
 as well as of the interaction
Lagrangians corresponds to that accepted in
[24], $g^2/4\pi=0.84$, $f/g=6.1$.
Note that $\alpha$ in (18) is the
 light-cone fraction of the proton momentum carried by the nucleon in the
$\rho N$ Fock state. The  applicability region of the above formula for
$f^T_{\rho N}(\alpha,\kt^2)$ is not much broader than
  $0.3\lsim \alpha\lsim 0.7$\,.

The data [2,3] on electroproduction
of single charged pions from hydrogen targets
were taken for $W>2.15\,GeV$ and
$\q2=0.6-9.8\, GeV^2$ with values of the polarization parameter
$$\epsilon={1-y-m_N^2Q^2/s^2\over 1-y+{1\over 2}y^2+m_N^2Q^2/s^2}$$
in the range $0.35<\epsilon<0.45$. Combination with data taken at
$\epsilon$ near 1 allowed the authors [2,3] to separate the contribution
from transversely polarized and longitudinal photons in the range
$1.2\,GeV^2<\q2<3.3\,GeV^2$.

Figure 2 shows the measured angular dependence of the transverse
$d\sigma_T/d\Omega_{\pi}$
 and
longitudinal
$d\sigma_L/d\Omega_{\pi}$
 components of the virtual photoproduction
 reaction $\gamma^*p\to \pi^+n$ for two ($W,\q2$) points.
The angle $\theta$ is the centre-of-mass angle between the pion
and the virtual photon.

For the electromagnetic pion form factor we use the expression
$F_{\pi}=(1+\q2/\Lambda_{\pi}^2)^{-1}$ with $\Lambda_{\pi}^2=m_{\rho}^2$.
 The so-obtained longitudinal (scalar)
cross section $d\sigma_L/d\Omega_{\pi}$ is shown in Fig.2. We conclude that
there is no dramatic difference between the data and the
 pion-pole
dominance model suggested long ago [1]. It is clear, however, that data
leave enough room for the variations of the cut-off parameter $\Lambda_{\pi}$
as well as for the extra non-$\pi$-pole contributions to $\sigma_L$ [5].

The
$\rho$-dominated transverse cross section is expected to be an
increasing function of $\theta$
in the range $\theta=0^0-20^0$. This effect  is due to the strong
$\kt$-dependence of the tensor term
$$
{f\over 4m_N}{\bar \psi}\sigma_{\mu\nu}{\psi}
(\partial^{\mu}{\rho}^{\nu}-\partial^{\nu}{\rho}^{\mu})
$$
 in the $\rho NN$ interaction
Lagrangian (terms $\propto f/g$, $\propto f^2/g^2$ in eq.(18)). Remind
that the ratio of the tensor to vector coupling constant, $f/g$, in
eq.(18) is large.

The measured angular dependence of the transverse component of the cross
section
 is in quantitative agreement with
our estimates at
 $\q2=3.3\,GeV^2$.

However, the above comparison suffers from some uncertainties.
First of all, the absolute normalization
depends strongly on the accuracy of determination
of the values $\q2$ and $W$
since
the $\rho N$ light-cone density (18) is a sharp function of
 the  variable
 $x\simeq\q2/(\q2+W^2)$
  in the range of $x\simeq 0.2-0.3$ corresponding
 to the above values of $\q2$ and $W$.  The resolutions obtained in [2,3]
are of about $0.5\, GeV^2$ in $\q2$ and $1\,GeV$ in $W$, not better.

Being taken at the face value, the data would imply the substantial excess of
$\rho$-mesons above the pions at $x\sim 0.2-0.3$. However, we keep the
ratio of the $\rho N$ and $\pi N$ light-cone densities at approximately
the same level
as in [11] to avoid contradiction with  data on the proton-neutron
charge exchange inclusive reactions, the universal cut-off parameter
in eq.(10) is put equal to $R^2=0.8\,GeV^{-2}$. Then, to reproduce the
observed $\q2$-dependence of the transverse cross section we have to assume
that the form factor $F_{\rho\pi}(\q2)$ has a very slow preasymptotics
and matches the pQCD behaviour (16) only at very high
$\q2\gg\Lambda^2_{\rho\pi}$.
Our estimates presented in Fig.2 correspond to
$\Lambda_{\rho\pi}=3m_{\rho}$. Evidently, the accurate measurements
 of the $x$-dependence
of the transverse component of the cross section must precede
any determinations  of the $\rho\pi\gamma$ form factor. Such an analysis
is also of interest in its own right, as it  provides
direct information on the $\rho N$ light-cone density function.

    The possibility to detect neutral mesons [25] makes the reaction
\beq
 ep\to e\pi^0 p
\endeq
particularly interesting as it is free of the pion-pole term and at high
$\q2$ is dominated by the transverse $\omega$-pole contribution.
Its differential cross section
reads
\beq
{d\sigma(ep\to e\pi^0 p)
\over dxdQ^2dk^2_{\perp}}=
K_T(x,\q2)f^T_{\omega N}(1-x,\kt^2)
{1\over 8}\q2
 F^2_{\omega\pi}(\q2)\,,
\endeq
where the density function $f^T_{\omega N}$ coincides with
$f^T_{\rho N}$ at $f/g=0$ and $g_{p\rho^0 p}\to g_{p\omega p}$ [24].

The coupling constant $g_{\omega\pi\gamma}$ is large. Experimentally [26],
\beq
{g^2_{\omega\pi\gamma}\over g^2_{\rho\pi\gamma}}\approx 10.\,,
\endeq
which is in agreement with the early $SU(3)$ quark model predictions
 [21,22]
\beq
F_{\rho^{\pm}\pi^{\pm}}=F_{\rho^0\pi^0}={1\over
3}F_{\omega\pi}\,.
\endeq
 Due to the extra power of $\q2$ typical of the
 magnetic dipole interaction the cross section of the single  $\pi^0$
production at high $\q2$ is dominated by  the $\omega-\pi^0$ radiative
   transition.  Evidently, it holds true even at the "normal" value of the
parameter $\Lambda_{\omega\pi}\simeq m_{\rho}\,$.  This observation makes
 the
 idea of the precision measurements of
the $\q2$-dependence of the cross section of
 the reaction $p(e,e^{\prime}\pi^0)p$ [25]
especially appealing.

The appearance of the
pQCD-generated longitudinal component of the cross section (21) is
discussed in [5].

Note, the term corresponding to the contribution of the diagram of Fig.1b
$$K_T(x,\q2)
 f_{\pi N}(x,k^2_{\perp})
F_{p}^2(\q2)\,,$$
where $F_p(Q^2)=F_1(Q^2)$ is the Dirac form factor of the proton
is of the next order in powers of $Q^{-2}$  and can be neglected.
There is also the kinematical suppression of the small-$x$ region, since
the function $f_{\pi N}(x,\q2)$ peaks at $x\simeq 0.7$ and steeply falls
at $x\to 0$.

Note, the non-zero contribution to
the cross section of the  reaction (20)
 could come from
the interference diagram of Fig. 1c\,. This is typically a non-partonic
 contribution.
In our particular case the
disbalanced  kinematics of the $\pi N$ system (see above)
generates  large invariant masses of the intermediate $\pi N$ state
 which are about
$$
\sim
{m_N\over m_{\pi}}
$$
times as large as the values of $M^2_{\pi N}$ typical of the diagrams 1a,b.
 This is the reason why the interference diagram
 (Fig.1c)
loses the competition with both the diagrams 1a and 1b
 for all $x$'s,
 independently of the specific form of
the  vertex function.

\section{Final-state interactions and the color transparency phenomenon}

In the reaction
$p(e,e^{\prime}\pi^+)n$
at  $\q2\sim 5-10\,GeV^2$
the cms energy squared of the pion-nucleon final state rescattering
 is large,
\beq
s_{\pi N}\simeq {1-x\over x}\q2
\endeq
and we can use the Glauber approximation [27,28,29] for the final-state
interaction (FSI) of the struck pion with the spectator neutron.
Then the pion-nucleon density, $f_{\pi N}(x,\kt^2)$,
which plays the role of the  momentum distribution of the observed
pion (nucleon) can be rewritten as
\beq
f^{FSI}_{\pi N}(x,\kt^2)=\left|\phi_{\pi N}(x,\vec\kt)-
{1\over 4\pi}\int d^2\vec \ks
\phi_{\pi N}(x,\vec\kt-\vec \ks)
f_{el}(\vec \ks)\right|^2,
\endeq
where
\beq
f_{el}(\vec \ks)=
{\sigma^{tot}_{\pi N}\over 4\pi}\exp
\left(-{1\over 2}b_{el}{\vec \ks}^2\right)
\endeq
is the $\pi n$ elastic scattering amplitude with the slope parameter
$b_{el}= 10 \,GeV^{-2}$ [30].

To integrate over $\vec\ks$ in (25) it is convenient to rewrite
 $\phi_{\pi N}$ as a product
\beq
\phi_{\pi N}=\chi\cdot\Gamma,
\endeq
where $\chi(x,\vec\kt)$ is a smooth function of $\kt$, while the most
singular part of $\phi_{\pi N}$ is the vertex function $\Gamma(M^2_{\pi N})$.
Then  factoring
  out of the integral
the slowly varying
 function $\chi(x,\vec\kt)$
 enables one to
write the momentum distribution, corrected for FSI effects,
$f^{FSI}_{\pi N},$ in a factorized form:
\beq
f^{FSI}_{\pi N}(x,\kt^2)=
f_{\pi N}(x,\kt^2)\left[1-\eta(x,\kt^2)\right]^2,
\endeq
where the attenuation function, $\eta(x,\kt^2)$, equals
\beq
\eta(x,\kt^2)=
{\sigma^{tot}_{\pi N}\over
16\pi\left[B(x)+{1\over 2}b_{el}\right]}\exp
\left[{B^2(x)\kt^2\over B(x)+{1\over 2}b_{el}}\right]
\endeq
and
\beq
B(x)={1\over 2}{R^2_{\pi N}\over x(1-x)}.
\endeq
A quick estimate yields for the attenuation effect
\beq
1-2\eta\approx 0.75\,.
\endeq
Still, the FSI effects distort the transverse momentum distribution.
Thus, an accurate evaluation of the FSI effects is necessary for the
quantitative interpretation of the data.

Eq.(25) implies that the attenuation is due to the interference of
the elastic rescattering amplitudes. It has been shown in [31] that at
high $Q^2$ the interference of elastic ($\pi$, in our case)
 and inelastic ($A_1$, for instance) intermediate states
gives rise to delicate cancellations of the
contributions from the ground state and higher excitations
thus resulting in CT phenomenon and/or weak final state interaction.

The patterns of the multiple scattering on both the nucleon and the nucleus
are similar ones. But there is one important difference.
A slow onset of CT on nuclei predicted in [31] and confirmed
by the recently completed NE18 experiment [32] is due to the
nuclear form factor suppression of the high mass intermediate states:
\beq
M^2_{max}\simeq
{s_{\pi N}\over m_NR_A}\,.
\endeq
Then to get the $Q^2$-dependence  of the strength of FSI denoted as
 $\Sigma_{\pi N}$,
 it suffices to  rescale
 the function $\Sigma_{\pi A}(\q2)$, already calculated in [34]
 making use of the diffraction operator technique, as follows
\beq
\Sigma_{\pi N}\left(Q^2,x\right)=
\Sigma_{\pi A}\left({R_A\over R_N}s_{\pi N}\right)\,.
\endeq
Then, a quick estimate of the CT effects comes from substituting
$\sigma^{tot}_{\pi N}$
 in eq.(29) by
$\Sigma_{\pi N}(Q^2,x)$.
 It can easily be seen  that already at $Q^2=10 \,GeV^2$
 and $x\sim 0.5$ the cross section $\Sigma_{\pi N}(Q^2)$
is
  approximately half
the "normal" $\pi N$ cross section at the energy $\simeq 10\,GeV $  and
vanishes at $Q^2\simeq 30\, GeV^2$. Note that the rescaling (33)
can not be quite
exact at $\q2\sim 3\,GeV^2$, near the threshold of the
first radial excitation ($A_1$-meson). Due to the large $\pi-A_1$ mass
splitting,
there is a region of very slow variations of
$\Sigma_{\pi N}(s_{\pi N})$ which is followed by a sharp fall off.

	This observation is important for the correct continuation of
 the vertex function $\Gamma(M^2)$, determined in the physical region
of $M^2\geq (m_B+m_M)^2$, to the meson-pole at $M^2=m_N^2$.
Indeed, any meson-baryon LCWF
  having a claim on description of the hadronic processes
involves implicitly the FSI effect.
 Disappearance of the latter at high $\q2$ implies that
at moderate $Q^2$ the effective  $\pi NN$ coupling constant
 does not equal its standard pole value and must be
renormalized.
The renormalization amounts to the $25\%$ effect.


\section{Conclusions}

We analyze the high-$Q^2$ exclusive electroproduction
in terms of the light-cone parton densities with mesons and baryons as
the (non-perturbative) partons.
In the partonic kinematics the observable cross sections
are expressible directly in terms of the on-shell
electromagnetic form factors.
The transverse component of the cross section $ep\to e\pi^+n$ is dominated
by the $\rho$-pole contribution. As a hint to the relevance of such an
 interpretation of  data we consider the observed angular distribution
of charged pions.

We consider the opportunity  of measuring
the magnetic dipole  form factors
$\rho\pi\gamma$, $\omega\pi\gamma$ in the  reaction with the
charged and neutral pions as a very  promising one [25]. In particular,
it has been emphasized above
that at high $\q2$ the helicity non conserving $\omega-\pi$ transition
prevails in the $p(e,e^{\prime}\pi^0)p$ reaction due to both the specific
$\q2$-enhancement of the $M1$ radiative transition and large value
of the $\omega\pi\gamma$ coupling.

Note, the early onset of the parton model regime  enables
to study the form factors $F_{\rho\pi}(\q2)$, $F_{\omega\pi}(\q2)$
 in the substantially non-perturbative region and, what is more important,
to retrace the onset of the pQCD regime at a very high $\q2$ as well.

\vspace{1cm}

{\bf Acknowledgments:} We benefited a lot from discussions with
N.N.Nikolaev.  Useful comments of S. Jeschonnek and A.
Bianconi are gratefully acknowledged.  This work was partially supported
by the INTAS grant 93-239 and G.Soros International Science Foundation
grant  N MT5000.

\newpage

{\bf \Large Figure captions:}
\begin{itemize}

\item[Fig.1]
- The one-meson electroproduction on the proton.
The virtual photons, mesons and baryons are represented by wavy,
dashed and solid lines, respectively.

\item[Fig.2]
- The observed angular dependence of the transverse and longitudinal
components of the cross section for the reaction $\gamma^*p\to \pi^+n$
for the two ($W,\q2$) points. The data points are from [3].
 Our estimates are shown by the solid curves.

\end{itemize}
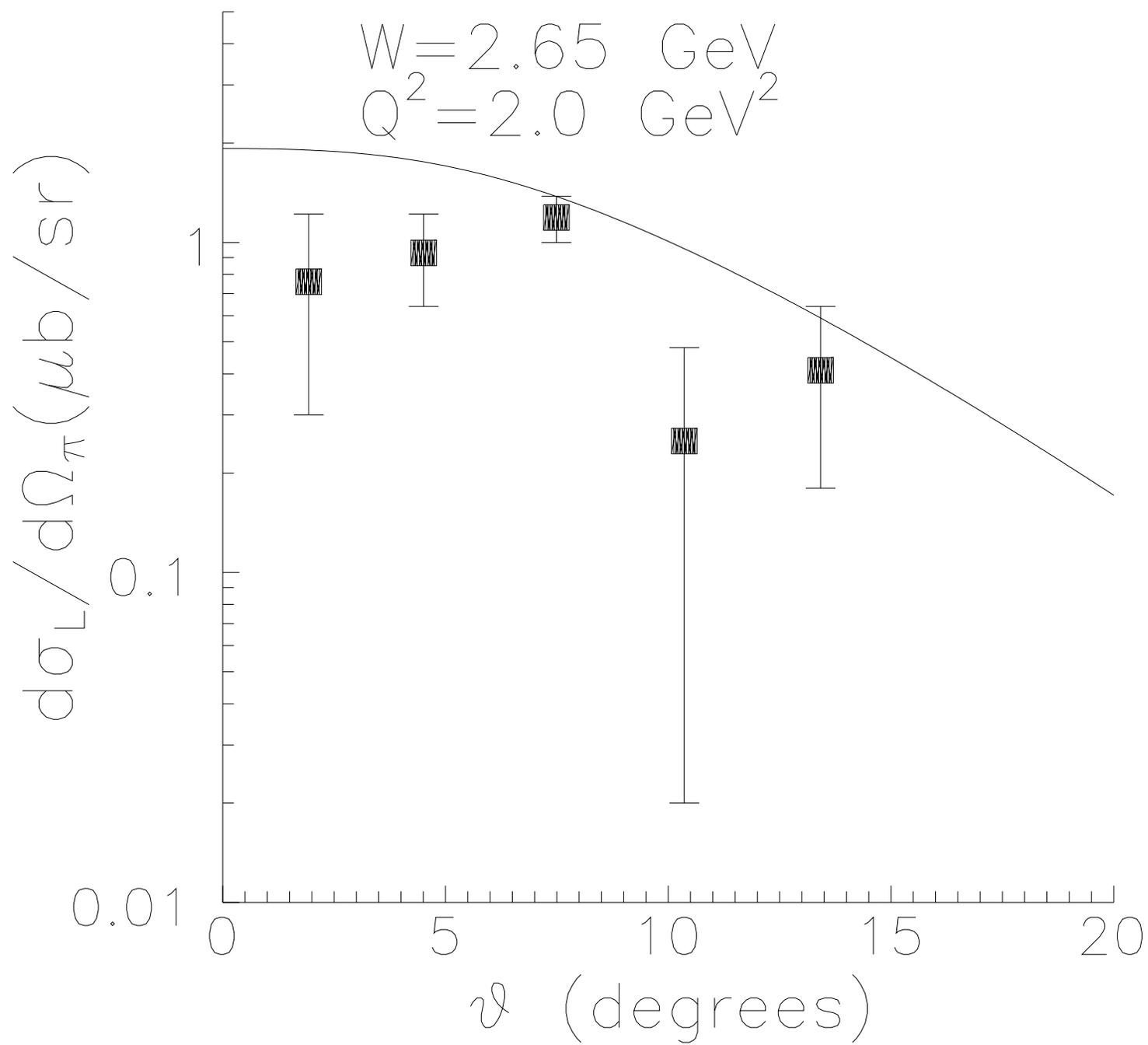
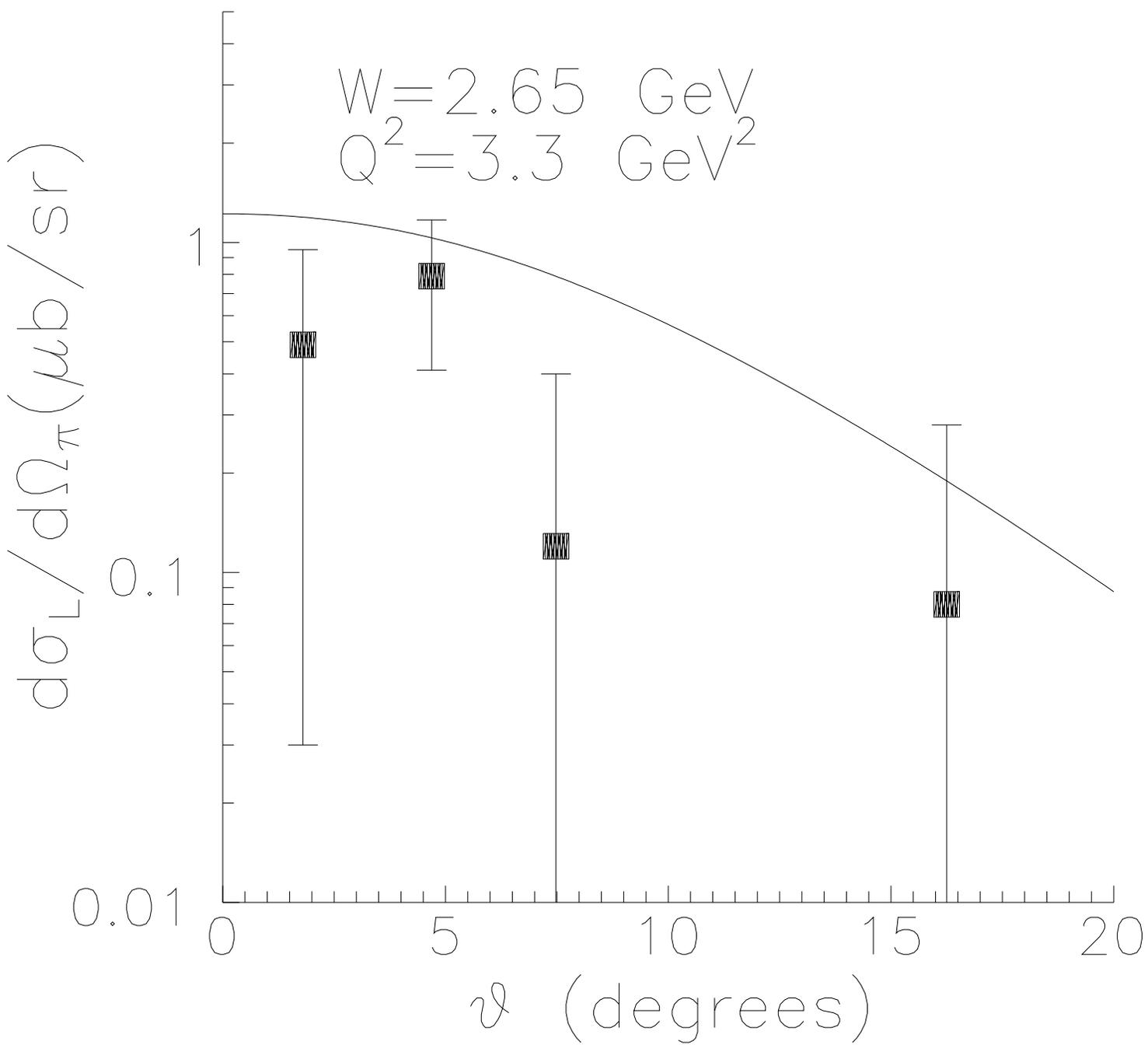
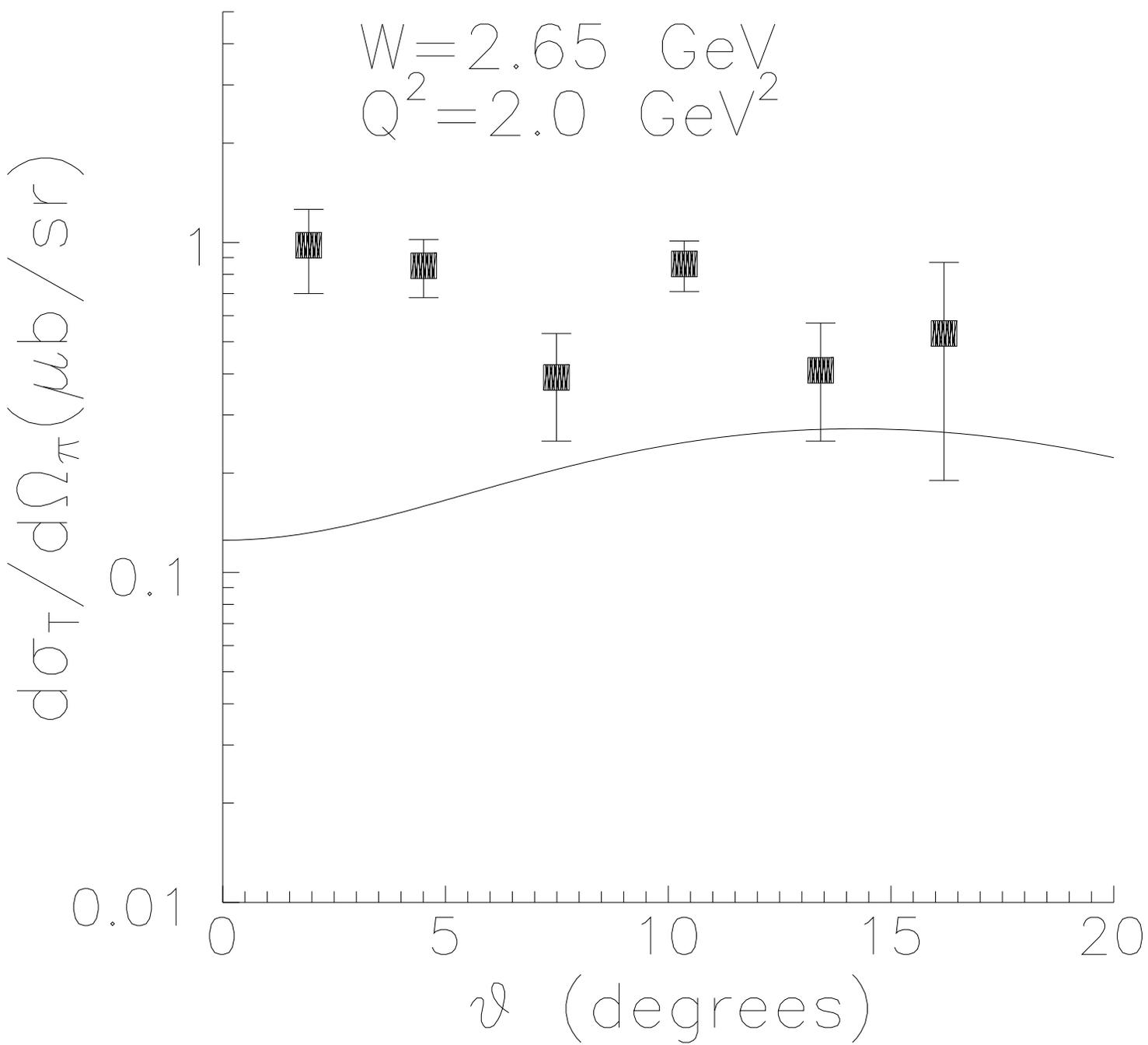
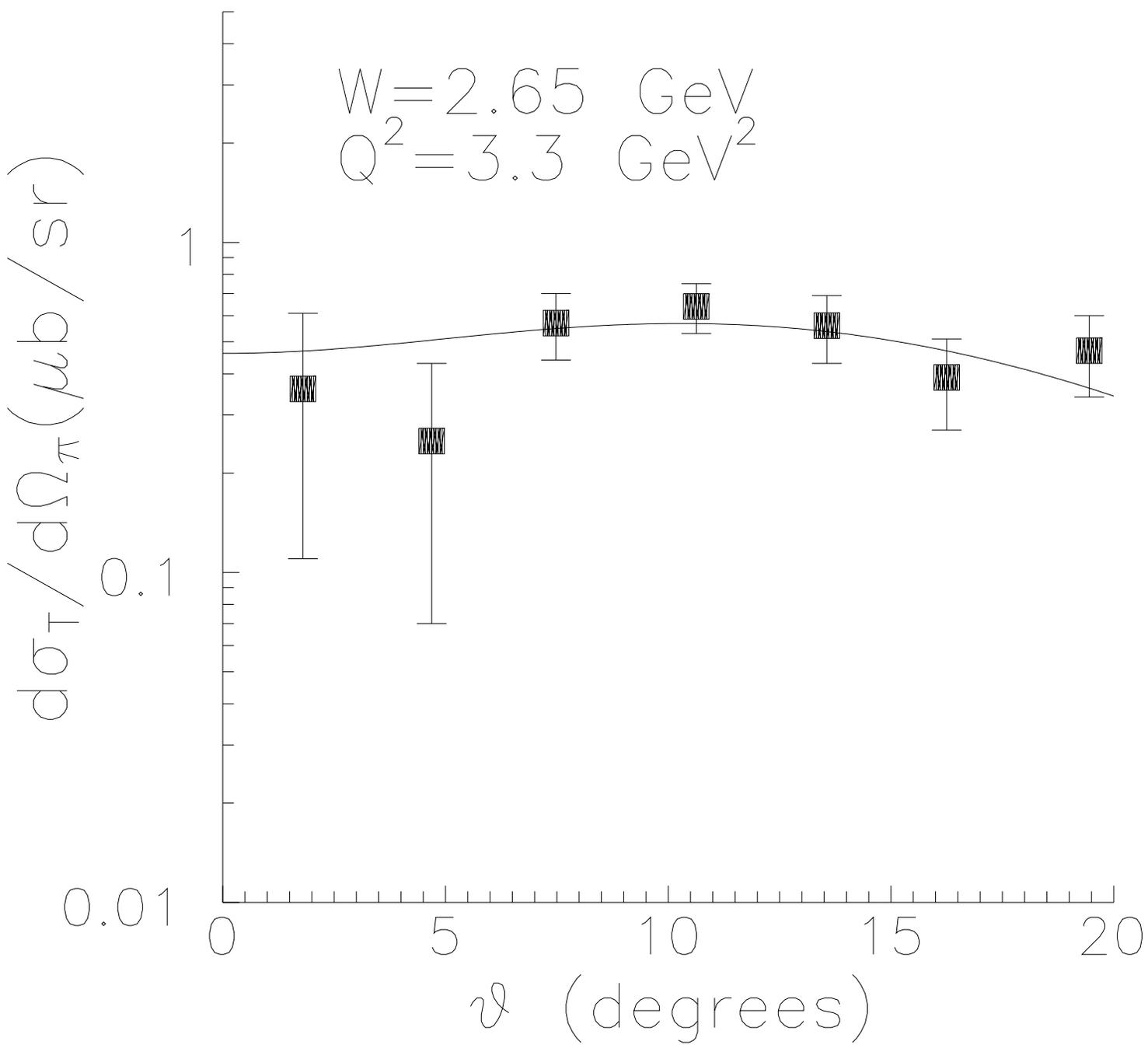
\end{document}